\documentclass[11pt,letterpaper]{article}

\usepackage[margin=1in]{geometry}
\usepackage{graphicx}
\usepackage{booktabs}
\usepackage{hyperref}
\usepackage{xcolor}
\usepackage{amsmath}
\usepackage{natbib}
\usepackage{caption}
\usepackage{subcaption}
\usepackage{float}
\usepackage{enumitem}
\usepackage{multicol}
\usepackage{microtype}
\usepackage{makecell}
\usepackage{multirow}
\usepackage{titlesec}
\usepackage{fancyhdr}
\usepackage{lastpage}
\usepackage{authblk}
\usepackage{abstract}
\usepackage{tikz}
\usepackage[section]{placeins} 
\newcommand{\cblock}[3]{%
  \mbox{%
  \protect\hspace{-1.5mm}%
  \protect\begin{tikzpicture}%
    \protect\node[minimum size=2.5mm,
          fill={rgb,255:red,#1;green,#2;blue,#3}] () {};%
  \protect\end{tikzpicture}%
  }%
}
\usetikzlibrary{patterns}
\newcommand{\chblock}[3]{%
  \mbox{%
  \protect\hspace{-1.5mm}%
  \protect\begin{tikzpicture}%
    \protect\fill[fill={rgb,255:red,#1;green,#2;blue,#3}] (0,0) rectangle (2.5mm,2.5mm);%
    \protect\fill[pattern=north west lines, pattern color=white!90!black] (0,0) rectangle (2.5mm,2.5mm);%
  \protect\end{tikzpicture}%
  }%
}
\newcommand{\chblockdark}[3]{%
\mbox{%
\protect\hspace{-1.5mm}%
\protect\begin{tikzpicture}%
  \protect\fill[fill={rgb,255:red,#1;green,#2;blue,#3}] (0,0) rectangle (2.5mm,2.5mm);%
  \protect\fill[pattern=north west lines, pattern color=black!90] (0,0) rectangle
(2.5mm,2.5mm);%
\protect\end{tikzpicture}%
}%
}
\newcommand{\ccircle}[3]{%
  \mbox{%
  \protect\hspace{-1.5mm}%
  \protect\begin{tikzpicture}%
    \protect\node[circle, minimum size=2.5mm,
          fill={rgb,255:red,#1;green,#2;blue,#3}] () {};%
  \protect\end{tikzpicture}%
  }%
}

\definecolor{atomred}{HTML}{DC2626}
\definecolor{atomblue}{HTML}{2563EB}
\definecolor{atomgray}{HTML}{6B7280}
\definecolor{linkblue}{HTML}{3C3B6E}

\hypersetup{
    colorlinks=true,
    linkcolor=linkblue,
    citecolor=linkblue,
    urlcolor=linkblue,
    pdfauthor={Nathan Lambert, Florian Brand},
    pdftitle={The ATOM Report: Measuring the Open Language Model Ecosystem},
}

\newcommand{\roundlogo}[1]{%
  \begin{tikzpicture}
    \clip[rounded corners=3.75pt] (0,0) rectangle (0.8cm,0.8cm);
    \node[anchor=south west,inner sep=0pt] at (0,0)
      {\includegraphics[width=0.8cm]{#1}};
  \end{tikzpicture}%
}
\newcommand{\footerlogos}{%
  \roundlogo{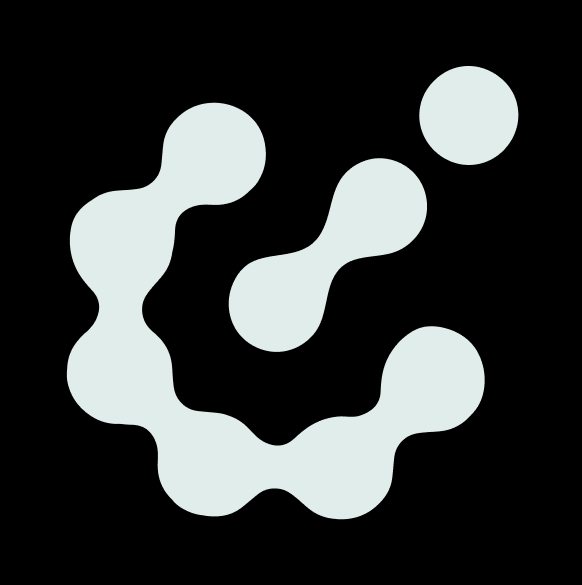}\hspace{3pt}\roundlogo{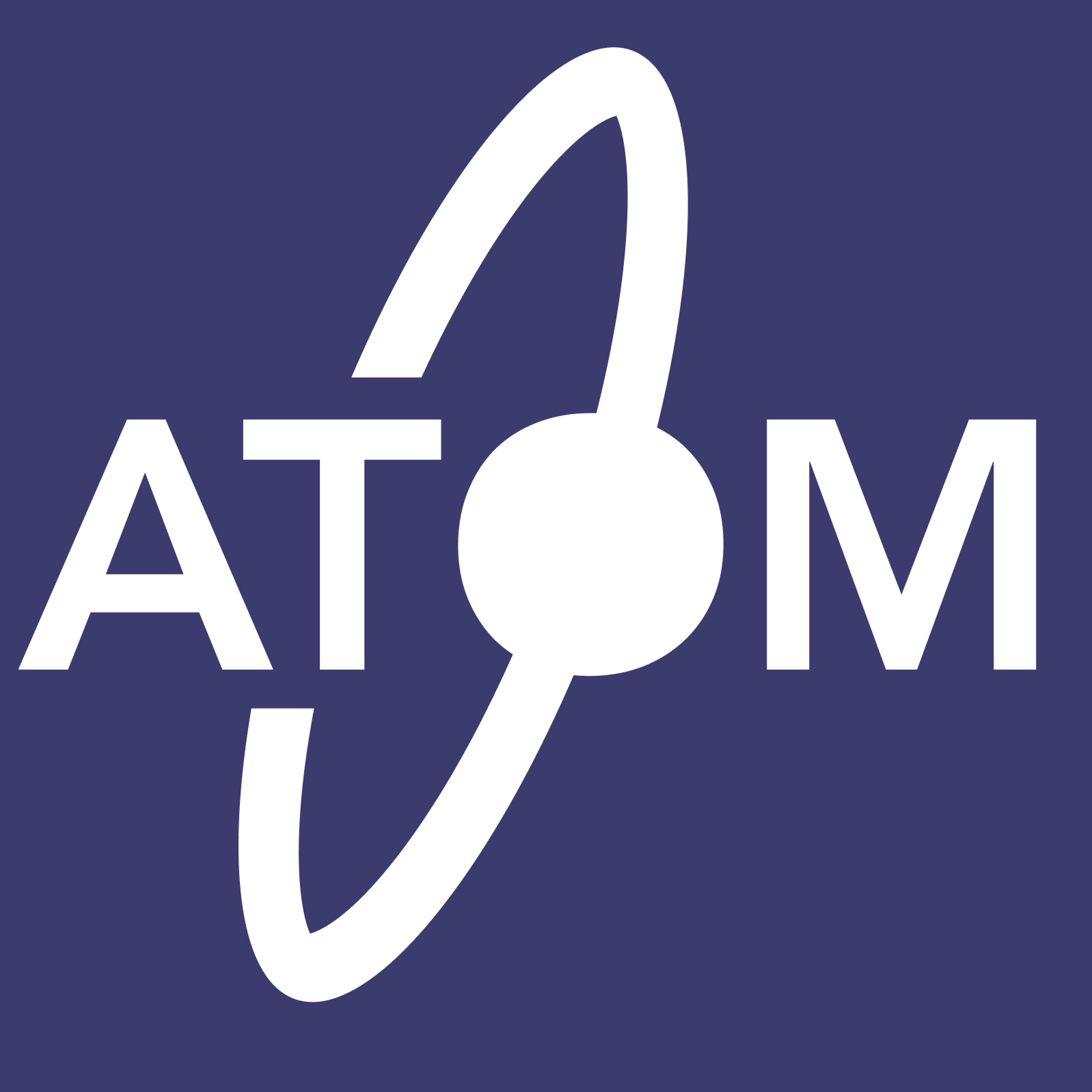}%
}

\titleformat{\section}{\large\bfseries}{\thesection}{1em}{}
\titleformat{\subsection}{\normalsize\bfseries}{\thesubsection}{1em}{}

\graphicspath{{figures/}}

\title{The ATOM Report:\\ Measuring the Open Language Model Ecosystem}
\author[1]{Nathan Lambert}
\author[1]{Florian Brand}
\affil[1]{\small{Interconnects AI, \href{https://atomproject.ai}{atomproject.ai}}}
\date{}

\begin{document}

\maketitle
\vspace{-2em}
\thispagestyle{empty}

\begin{abstract}
\vspace{-9pt}
\noindent
We present a comprehensive adoption snapshot of the leading open language models and who is building them, focusing on the $\sim$1.5K mainline open models from the likes of Alibaba's Qwen, DeepSeek, Meta's Llama, that are the foundation of an ecosystem crucial to researchers, entrepreneurs, and policy advisors.
We document a clear trend where Chinese models overtook their counterparts built in the U.S. in the summer of 2025 and subsequently widened the gap over their western counterparts.
We study a mix of Hugging Face downloads and model derivatives, inference market share, performance metrics and more to make a comprehensive picture of the ecosystem.
\end{abstract}

\begin{figure}[h]
\vspace{-6pt}
  \centering
  \includegraphics[width=\linewidth]{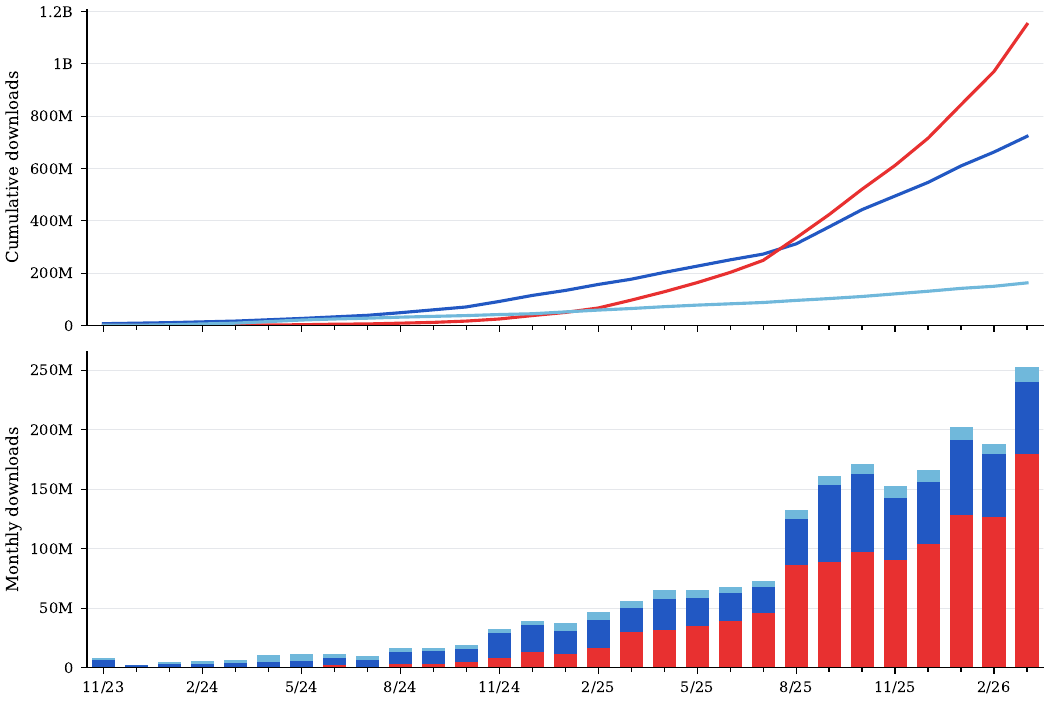}\\
  \cblock{232}{48}{48} China \quad \cblock{112}{184}{219} EU \quad \cblock{34}{88}{195} USA
  \caption{Cumulative open model downloads by region, November 2023 -- March 2026. Chinese models overtook American models by August 2025, with the gap widening to $>$400M downloads.}
  \label{fig:global_momentum}
\end{figure}

\newpage
\setcounter{tocdepth}{2}
\tableofcontents
\newpage

\section{Introduction}
\label{sec:introduction}



Open weight AI models are becoming foundational infrastructure across research, startups, and governments negotiating their future in understanding, building, and deploying increasingly powerful AI systems.
During this time period of rapid advancements in AI capabilities and training methodologies, the role of open models is constantly evolving through different families of models and performance gaps to the best, closed frontier models.
Within all of this, measuring the \textit{adoption} of open models in parallel to the substantial work on model performance and quality (including Arena, formerly ChatBotArena~\citep{chiang2024chatbotarena}, Artificial Analysis's Intelligence Indices~\citep{artificialanalysis2026intelligencebenchmarking}, and Epoch AI's benchmarking~\citep{epochai2025eci}), is often opaque and hard to draw insights from.
In this report we detail our findings across different data sources to show the current key trends across open model usage.

The data for this report is expanded from the initial data methods used to make the case for The ATOM Project~\citep{lambert_atom_project_2026}, and it is designed to inspire more precise investment and innovation on open language models.
This report does not make specific policy recommendations, for those, see The ATOM Project essay at \url{https://atomproject.ai/}.
A key function of The ATOM Report and its broader related initiatives\footnote{Such as extended analysis at \url{interconnects.ai}. Examples are \href{https://www.interconnects.ai/p/8-plots-that-explain-the-state-of}{here} or \href{https://www.interconnects.ai/p/2025-open-models-year-in-review}{here}.}, is to focus on the specific question of the development of \textit{open language models} specifically and not all of open-source AI collectively.
This involves curating a specific subset of the public data on open models and curating the rough group of models that is defining adoption patterns -- e.g. many small, specific open models such as classifiers models can dominate download or fine-tuning numbers, while having minimal impact on technological advancements.

We detail the following key findings in this report:
\begin{enumerate}
    \item \textbf{Chinese open models moved from trailing the U.S. to a clear lead in cumulative adoption}: China overtook the U.S. in late July 2025 and reached 1.15B cumulative downloads versus 723M by March 2026, as shown in Fig.~\ref{fig:global_momentum}.
    \item \textbf{Qwen is the single most-used open model family overall}: The growth of the Chinese model ecosystem can largely be attributed to Alibaba's Qwen, which is responsible for almost a billion cumulative downloads by March 2026. Other open model families such as Llama, DeepSeek or Kimi lag far behind. The growth of Qwen-based models has even accelerated especially with the updated model series of Qwen3 2507 and Qwen3.5, respectively.
    \item \textbf{Other adoption metrics back up the trend}: Inference data from OpenRouter shows Chinese models taking a large lead in the summer of 2025, with the gap increasing drastically towards the end of the year. This is largely correlated with Hugging Face downloads and model derivatives. DeepSeek leads in OpenRouter measurements, where the total usage is more split across organizations relative to Hugging Face metrics.
    \item \textbf{We introduce the Relative Adoption Metric (RAM)}: To showcase the momentum of model adoption beyond just raw download numbers, which are primarily dominated by initial download numbers for small models, we develop a relative metric over time and model size. We find that, among large general models, GPT-OSS 120B and Nemotron 3 Super show exceptional adoption at roughly 17--20$\times$ their size-class target, while DeepSeek V3.2 and GLM 4.7 underperformed at well below 1$\times$ their respective baselines. This suggests the relative popularity, even long after release, of US-based models.
\end{enumerate}

\section{Methodology}
\label{sec:methodology}

\subsection{Data Sources and Collection}

We rely on a variety of data sources in this report.
The primary data is composed of aggregated, public data on the open model platform Hugging Face, which measures downloads of each model and highlights model derivatives (i.e. which models are used as base models for fine-tuning).
We also include data from public benchmarking websites Artificial Analysis~\citep{artificialanalysis2026intelligencebenchmarking} and Arena (formerly ChatBotArena)~\citep{chiang2024chatbotarena} to showcase model performance and other adoption data from the popular open model inference platform \url{OpenRouter.ai}.

\paragraph{Hugging Face Downloads.}
The primary adoption signal is cumulative download counts over time from the Hugging Face Hub, which is the primary destination for open models and provides publicly accessible data.
By taking daily snapshots of the total downloads of every model, we can showcase trends of data not directly on \url{huggingface.co}.
A ``download'' is defined as any HTTP request to the model file hosting endpoint, including programmatic access via \texttt{transformers}, \texttt{curl}, and similar tools\footnote{See the \href{https://huggingface.co/docs/hub/en/models-download-stats\#how-are-downloads-counted-for-models}{Hugging Face documentation} for more details}.
We obtained historical monthly download data directly from the Hugging Face team for the leading open model organizations -- Meta Llama, Qwen, Mistral AI, DeepSeek, Google, and Microsoft -- covering November 2023 through July 10, 2025.
Beginning in early July 2025, we operate an independent daily scraper that records total cumulative downloads and other public metadata for all publicly available models.
The historical series (through July~10, 2025) uses $2.5\times$ IQR outlier-filtered data (described below). From August 2025 onward, we compute monthly deltas from the unfiltered scraper data and add them to the filtered baseline, preserving continuity at the splice point without re-filtering the full history. Throughout, monthly labels refer to data as of the first of that month: ``Aug~25'' denotes cumulative downloads through July~31, 2025.

Throughout the report we refer to open models, open language models, and open-weight language models. The final term is the most precise, as we are measuring the adoption of any set of model weights available on the Hugging Face platform, regardless of license or information available about it (which could earn it the classification ``open-source'').

We track every prominent open language model released since ChatGPT -- i.e.\ models that accept text (and optionally other modalities) as input and produce text as output -- while excluding embedding models, text-to-image diffusion models, and other non-generative architectures.
The full list of $\sim$1.5K tracked models is publicly available.\footnote{\url{https://github.com/Interconnects-AI/tracked-models/blob/main/models.csv}}
In total, these models account for over 3 billion downloads across the study period (November 2023 through March 2026).
To mitigate noise on the initial data, we applied $2.5\times$ interquartile range outlier detection on daily download series; models without anomalous spikes are left unfiltered.
This substantially improved the historical data, which is prone to large variations and unexplainable features.

For derivative and fine-tune tracking, we identify derivatives via the Hugging Face \texttt{base\_model} tag, which records the parent model in the format \texttt{base\_model:ORG/MODEL}. We restrict to derivatives whose base model appears in our tracked model list, require more than five lifetime downloads, and exclude local-inference re-uploads (e.g., GGUF, MLX).

\paragraph{OpenRouter.}
Inference token share data is drawn from OpenRouter, which provided historical data on the top 10 open models per month ranked by total tokens served.
Only the top 10 are reported for each month, so organizations whose usage is spread across time or many models may be undercounted.
This provides a complementary demand signal, capturing which models people use, which is not reflected in download counts alone.

\paragraph{Arena.}
Aside from quantitative metrics, we also take into account qualitative metrics to showcase the progress of open models. Community Elo ratings come from Arena (formerly ChatBotArena), which measures general chat quality through blind pairwise comparisons by human evaluators~\citep{chiang2024chatbotarena}. We use the style-controlled overall Elo as the primary performance metric. Scores prior to May 19, 2025 are shifted upward by 59.2 points to account for a platform-wide recalibration when style control was set as default~\citep{li2024stylematter}. For each region, we log the highest-scoring open model at each snapshot date to showcase the regional performance frontier; scores are enforced to be monotonically non-decreasing (despite scores fluctuating slightly due to the statistical variance in Arena's platform) so that the frontier only advances.

\paragraph{Artificial Analysis.}
The Overall Intelligence Index from Artificial Analysis aggregates scores across standard academic benchmarks for leading language models. We extract the top open model per region over time and fit a simple linear regression to highlight performance trends, beginning from models released after April 2024.

\subsection{Model Categorization}
\label{sec:categorization}

\begin{figure}[H]
  \centering
  \includegraphics[width=\linewidth]{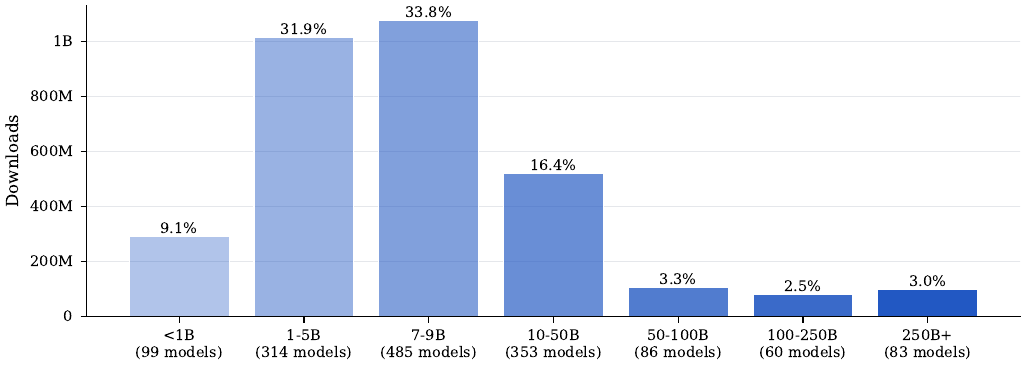}
  \caption{Distribution of our tracked open model downloads by parameter count.
  The 7--9B range captures 33.8\% of all downloads, with sub-10B models accounting for $\sim$75\%.}
  \label{fig:downloads_by_size}
\end{figure}

\paragraph{Size Buckets.}
As size is one of the most important factors to select the right model, we partition models into seven parameter-count categories: ${<}1$B, 1--5B, 7--9B, 10--50B, 50--100B, 100--250B, and 250B+.
We seperate the 7--9B range from the broader 1--10B span because it captures a natural concentration of popular model architectures (Llama~8B, Mistral~7B, Qwen~7B). This is also apparent in the data (see  Figure~\ref{fig:downloads_by_size}), where this bucket alone accounts for one third of all downloads, making it the single largest bucket, followed closely by the 1--5B bucket.
The remaining buckets represent natural partitions for models which have emerged over time, e.g. there have been standard model sizes of ~32B and 70B parameters for multiple years in the ecosystem due to popular hardware configurations.
For MoE models, we use total parameter count rather than active parameters per token---e.g., DeepSeek-R1 (671B total, 37B active) is placed in the 250B+ bucket.
Finer-grained analysis of MoE model sizes and sparsity relative to dense models is left to future work.

\paragraph{Regional Classification.}
Models are attributed to one of three regions---United States, China, or Europe---based on the headquarters of the releasing organization. This is an imperfect proxy, as it does not capture the nationality of individual contributors or even their organizations. The mapping is as follows:

\begin{itemize}[nosep,leftmargin=1.5em]
  \item \textbf{United States:} Meta (Llama), Google (Gemma), Microsoft (Phi), NVIDIA (Nemotron), OpenAI (GPT-OSS), Allen AI (Olmo), IBM (Granite), Snowflake, AI21 Labs
  \item \textbf{China:} Alibaba (Qwen), DeepSeek, ByteDance, Baidu, InternLM, Zhipu AI (GLM/ChatGLM), OpenBMB, Inclusion AI, Skywork, Tencent, Xiaomi (MiMo), Moonshot AI (Kimi), MiniMax
  \item \textbf{Europe:} Mistral AI, Hugging Face (SmolLM)
\end{itemize}

\subsection{Limitations}

Hugging Face is not the sole distribution channel for open models, and meaningful traffic can flow through other platforms such as ModelScope. 
Large deployments of open models can also count as just one download if a user downloads the model to local infrastructure and then never relies on the public versions. 
The latter is also the case for other services, such as clouds or other SaaS products, which might cache model weights for production usage. 
These factors might result in under reporting the popularity of certain model (families). 
While this decentralization makes it impossible to pinpoint exact numbers, the larger trends and the relation of the models to each others largely hold outside of this tracked usage.
The exclusion of re-packaged models in GGUF or MLX format from derivative counts avoids inflating fine-tune metrics with local-inference re-uploads, which are especially prevalent for small, popular models.

As for the metrics for usage, download counts do not equate to active usage, as automated CI/CD pipelines, bots, and repeated pulls inflate total downloads numbers, particularly for small models.  This effect might be stronger for models with "non-standard" architectures, as (small) models of a new architecture might be used in pipelines and tests to represent all current (and future) models of the same architecture.

However, we find that all available usage data is strongly correlated with the imperfect Hugging Face download metrics and show the same trend(s) and similar distribution.

\section{Model Adoption by Region}
\label{sec:global_landscape}



The core measurement of this report is following the balance of influence internationally among open model builders.
The overall rate and cumulative total of model downloads, binned per nation/region among the three largest contributors to open models -- The United States, China, and Europe generally -- is shown in Fig.~\ref{fig:global_momentum}.

\begin{figure}[H]
  \centering
  \includegraphics[width=\linewidth]{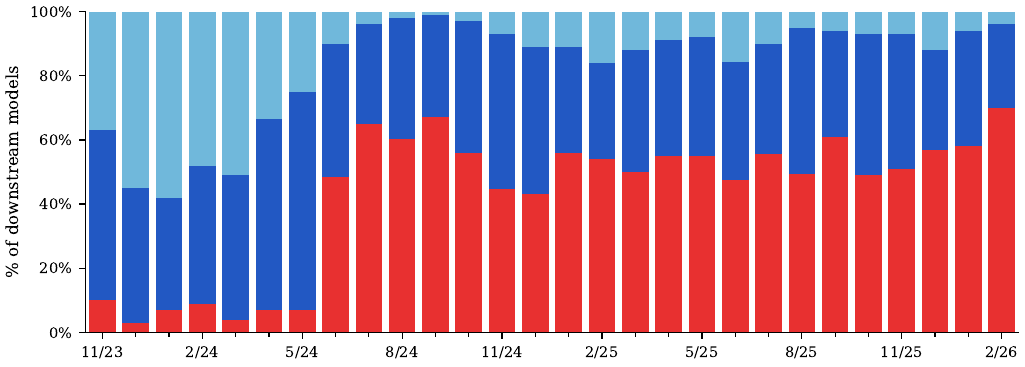}\\
  \cblock{232}{48}{48} China \quad \cblock{112}{184}{219} EU \quad \cblock{34}{88}{195} USA
  \caption{Share of new model derivatives by region per month. EU share fell from 58\% in January 2024 to 4\% by February 2026, while China rose to 70\%.}
  \label{fig:adoption_by_region}
\end{figure}

We observe three different phases:
After the release of ChatGPT, the EU dominated in terms of model adoption, exclusively driven by Mistral 7B \citep{jiang2023mistral7b} and Mixtral 8x7B \citep{jiang2024mixtral}. This was followed by a dominance of US-based models with the release of Llama 3 \citep{grattafiori2024llama3herd}, which has seen a lot of adoption due to its permissive license and the range of different sizes. Following the release of DeepSeek V3 \citep{deepseekai2024deepseekv3}, R1 \citep{deepseekai2025deepseekr1} and Qwen3 \citep{qwenteam2025qwen3}, the lead in usage went towards Chinese model builders, which have increased the gap since then.
In each of these eras, the respective country was responsible for \textgreater 50\% of downloads and fine-tuned models (see Figure~\ref{fig:global_momentum} for downloads and Figure~\ref{fig:adoption_by_region} for derivatives).
Cumulative tracked downloads reached 2.04 billion by March 2026, a 6$\times$ year-over-year increase from 339M in March 2025, highlighting the explosion of growth in open models.
China grew 11.9$\times$ YoY (97M to 1.15B) versus 4.1$\times$ for the USA (177M to 723M) and 2.5$\times$ for the EU (65M to 163M), widening the total download gap from 23M at August 2025 to 428M by March 2026.

The share of new model derivatives -- fine-tunes and adaptations built on base models -- follows the same pattern as downloads: China rose from 10\% in November 2023 to 70\% by February 2026, while the EU fell from a peak of 58\% to 4\% over the same period.

\begin{figure}[H]
  \centering
  \includegraphics[width=\linewidth]{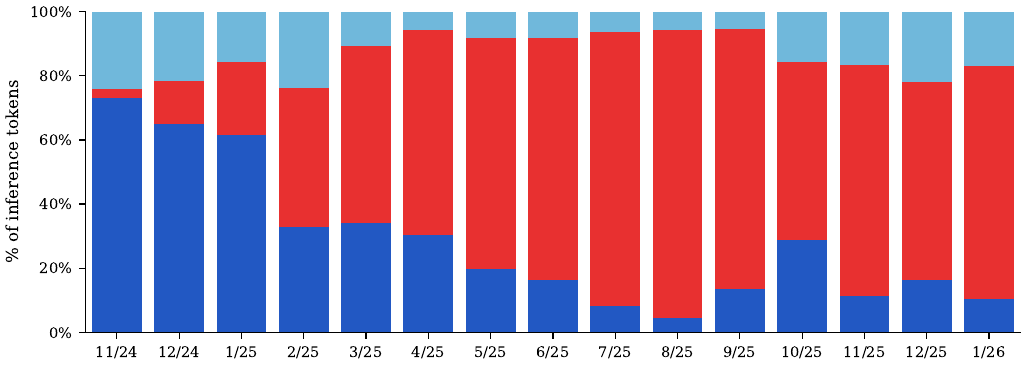}\\
  \cblock{232}{48}{48} China \quad \cblock{112}{184}{219} EU \quad \cblock{34}{88}{195} USA
  \caption{Open model inference token share by region (OpenRouter). China and the US completely inverted positions in 14 months, with China reaching 72.7\% by January 2026.}
  \label{fig:token_share_region}
\end{figure}

Open model inference data from OpenRouter shows the same shift with an even bigger magnitude (Figure~\ref{fig:token_share_region}): Chinese models' token share rose from 2.8\% to over 70\% in 14 months, while US-based models have shrunken considerably.

\clearpage
\section{Model Performance By Region}
\label{sec:performance}


To track the qualitative performance of leading open models over time, we rely on existing public benchmarking projects such as Arena~\citep{chiang2024chatbotarena} and Artificial Analysis's Intelligence Index~\citep{artificialanalysis2026intelligencebenchmarking}. For this, we took the open model subset of these leaderboards over time corresponding and plotted them per region with the categorization from Section~\ref{sec:categorization}.
Both benchmarks show a similar trend to the quantitative metrics, with the leading Chinese models overtaking the \textit{capabilities} of the leading American models late in 2024, holding a lead since. However the lead is less profound as in the quantitative metrics, which also makes sense -- a small lead in model performance can create a large lead in adoption, as uses of open models by default opt for the best model.

\begin{figure}[H]
  \centering
  \includegraphics[width=0.9\linewidth]{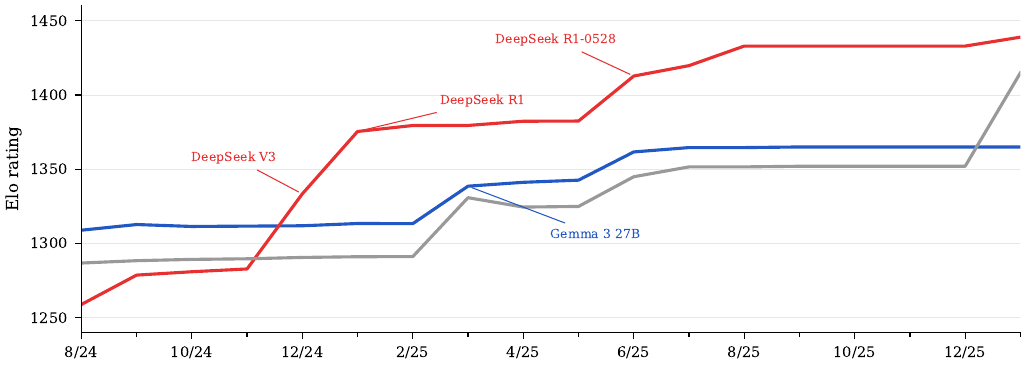}\\
  \cblock{232}{48}{48} China \quad \cblock{153}{153}{153} Other \quad \cblock{34}{88}{195} USA
  \caption{Arena Elo ratings for top open models by region. China surpassed the US in December 2024, driven by DeepSeek V3, and extended its lead through January 2026.}
  \label{fig:elo_rankings}
\end{figure}

\begin{figure}[H]
  \centering
  \includegraphics[width=0.9\linewidth]{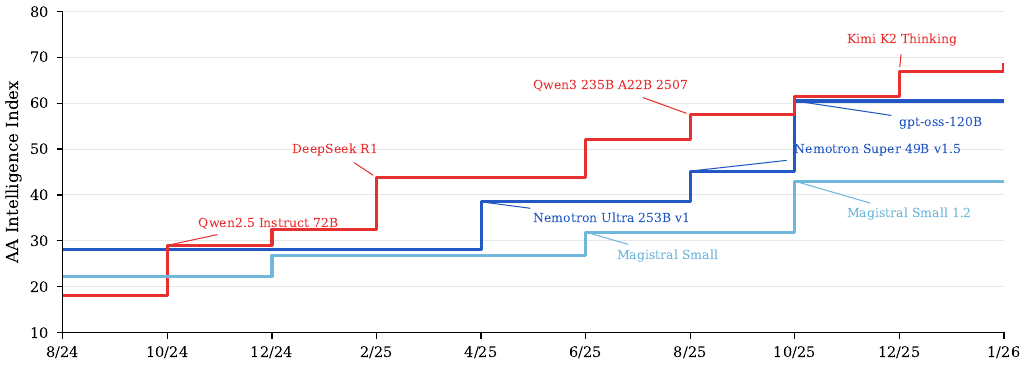}\\
  \cblock{232}{48}{48} China \quad \cblock{112}{184}{219} EU \quad \cblock{34}{88}{195} USA
  \caption{Artificial Analysis Overall Intelligence Index by region. China's best open model score quadrupled in 18 months, reaching 68.4 by January 2026.}
  \label{fig:best_models}
\end{figure}

The evolution of these benchmarks is shown for Arena's overall data in Fig.~\ref{fig:elo_rankings} and for Artificial Analysis's Index in Fig.~\ref{fig:best_models}.

\clearpage
\section{Model Adoption by Organization}
\label{sec:adoption}

To complement our analysis of open model performance and adoption by region, we also present the data split into individual organizations.
We balance the analysis between the evolution of the leading open model organizations, tracking the rise of Qwen, with the latest, emerging labs releasing excellent open models to challenge the incumbents.

\subsection{Ecosystem Leaders (Qwen, Llama, DeepSeek, Mistral, and OpenAI)}

\begin{figure}[b!]
  \centering
  \includegraphics[width=\linewidth]{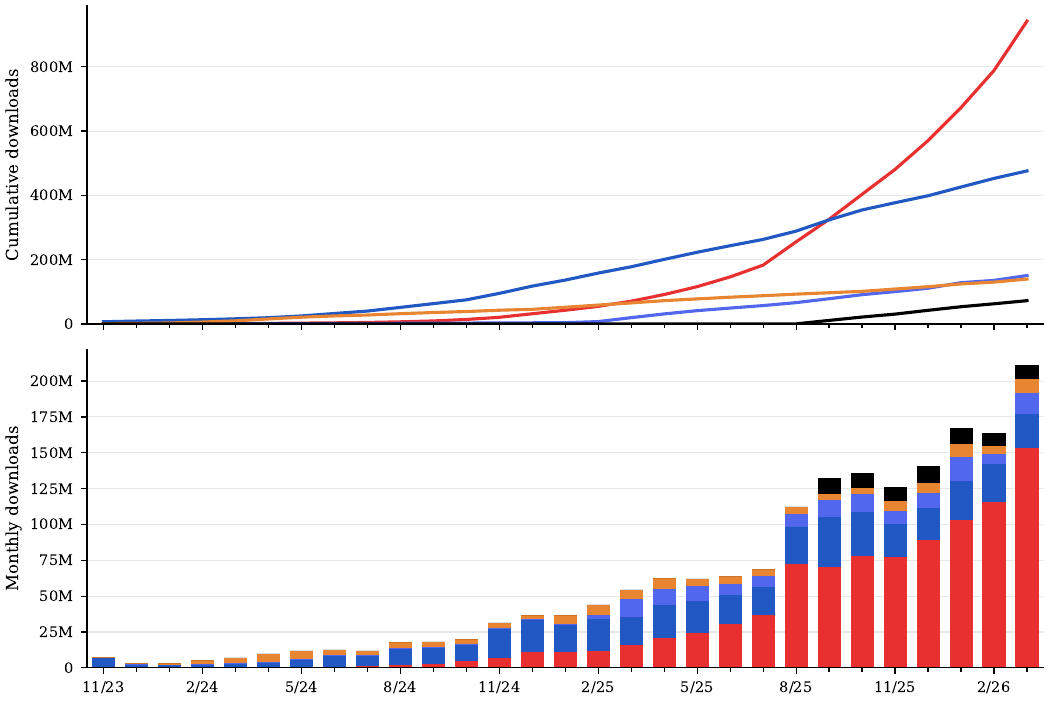}\\
  \cblock{81}{103}{237} DeepSeek \quad \cblock{34}{88}{195} Meta \quad \cblock{232}{133}{50} Mistral \quad \cblock{0}{0}{0} OpenAI \quad \cblock{232}{48}{48} Qwen
  \caption{Cumulative downloads for leading open model families. Qwen surpassed Llama in September 2025 and reached 942.1M downloads by March 2026.}
  \label{fig:china_momentum}
\end{figure}

\begin{figure}[H]
  \centering
  \includegraphics[width=\linewidth]{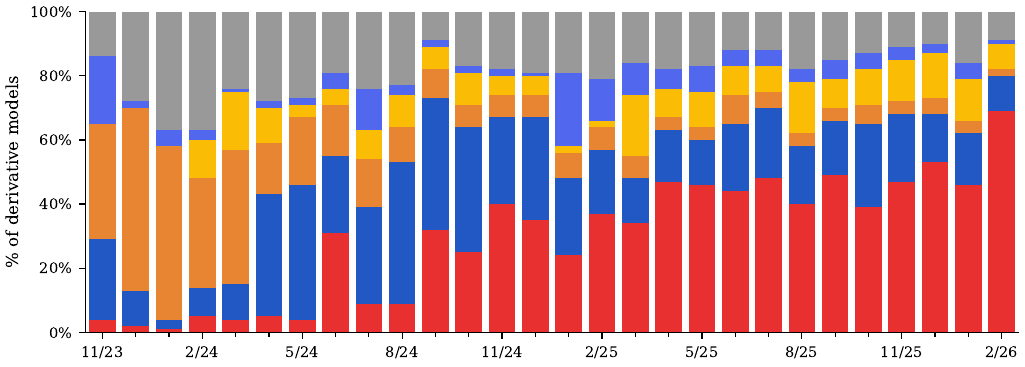}\\
  \cblock{81}{103}{237} DeepSeek \quad \cblock{251}{188}{5} Google \quad \cblock{34}{88}{195} Meta \quad \cblock{232}{133}{50} Mistral \quad \cblock{153}{153}{153} Others \quad \cblock{232}{48}{48} Qwen
  \caption{Monthly share of new model derivatives by organization. Qwen's derivative share reached 69\% by February 2026, while Meta's Llama fell from 25\% in November 2023 to 11\%.}
  \label{fig:adoption_trends}
\end{figure}

Figure~\ref{fig:china_momentum} shows cumulative downloads for five representative model families: Alibaba's Qwen, Meta's Llama, Mistral, DeepSeek, and OpenAI. 
Qwen surpassed Llama in cumulative downloads in September 2025 (325.4M vs.\ 323.7M) and by March 2026 reached 942.1M, thus nearly doubling the downloads of the Llama models (476.0M).
This change is even more stark in terms of derivatives, i.e., fine-tunes and other adaptations (such as LoRA adapters), where Qwen became the primary choice as a base model as early as June 2024.
Qwen's share of new fine-tunes and adaptations rose from 1\% in January 2024 to 69\% by February 2026 (Figure~\ref{fig:adoption_trends}), while Meta's peaked at 44\% in August 2024 before falling to 11\%.

This data also shows that new entrants can have impactful adoption, e.g. with OpenAI's new GPT-OSS models accumulating more adoption than long-established open model organizations such as Mistral AI.

\paragraph{Qwen's path to leading the ecosystem.}
The data of both model derivatives and downloads over time show how the ecosystem has evolved through multiple eras of model defaults.
The first two iterations of Meta's Llamas and Mistral's early models were the first platforms of the ecosystem through most of 2023.
Later, Meta's initial Llama 3 models were released in April of 2024, with the more popular 3.1 variants coming in July -- spurring Meta to an all-time lead in model derivatives in the second half of 2024.
These models carried Meta to an all-time downloads lead in Q1 of 2025, but Qwen was accelerating its rate of adoption.
A pivotal moment was the release of Qwen 2.5~\citep{qwenteam2024qwen25} in September of 2024, which has started the rise of Qwen in derivative models, which took substantial market share from Meta's Llama models and Mistral's early successful models (7B Dense and Mixtral 8x7B). 
The outlier in June 2024 is mostly due to spam, which persisted through our filtering of models with at least 5 life-time downloads.

Other prominent releases across the ecosystem, from Google's Gemma 2~\citep{googledeepmind2024gemma2} in July of 2024, Gemma 3~\citep{googledeepmind2025gemma3} in March of 2025, and many continued Mistral releases left their adoption metrics stagnant relative to Qwen's growth.
Qwen 3 in April 2025~\citep{qwenteam2025qwen3} (the same month as Llama 4's release) continued Qwen's acceleration. 
The early adoption numbers of Qwen3.5~\citep{qwen2026qwen35} since its release in February 2026 are an indication that the dominance of Qwen relative to its peers will continue (see Section~\ref{sec:ram-method} and specifically Fig.~\ref{fig:ram_case_studies} for early data on Qwen 3.5's adoption).

Over time, an equilibrium from May 2025 through March 2026 settled with Qwen having a base of 40\% or more of the derivative share, growing slowly, and the remainder being split predominantly between Meta, Google, Mistral, DeepSeek, and a long-tail of smaller labs.
Despite few model releases over the last year, Meta's Llama and Google's Gemma models maintained about 10\% share of derivative models uploaded to Hugging Face each, substantially more than any others.

\begin{figure}[H]
  \centering
  \includegraphics[width=\linewidth]{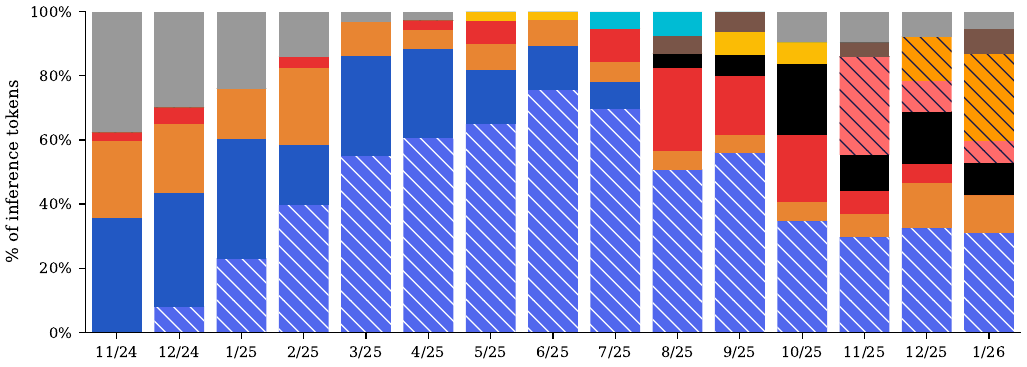}\\
  \chblock{81}{103}{237} DeepSeek \quad \cblock{251}{188}{5} Google \quad \cblock{34}{88}{195} Meta \quad \chblockdark{255}{107}{107} MiniMax \quad \cblock{232}{133}{50} Mistral \quad \cblock{0}{188}{212} Moonshot \\
  \cblock{0}{0}{0} OpenAI \quad \cblock{153}{153}{153} Other \quad \cblock{232}{48}{48} Qwen \quad \chblockdark{255}{152}{0} Xiaomi \quad \cblock{121}{85}{72} Zhipu
  \caption{Open model inference token share by organization (OpenRouter). Meta fell from a 37.4\% peak in January 2025 to zero, replaced by DeepSeek (31.1\%) and Xiaomi (27.2\%) by January 2026.}
  \label{fig:token_share_org}
\end{figure}

\begin{figure}[H]
  \centering
  \includegraphics[width=0.9\linewidth]{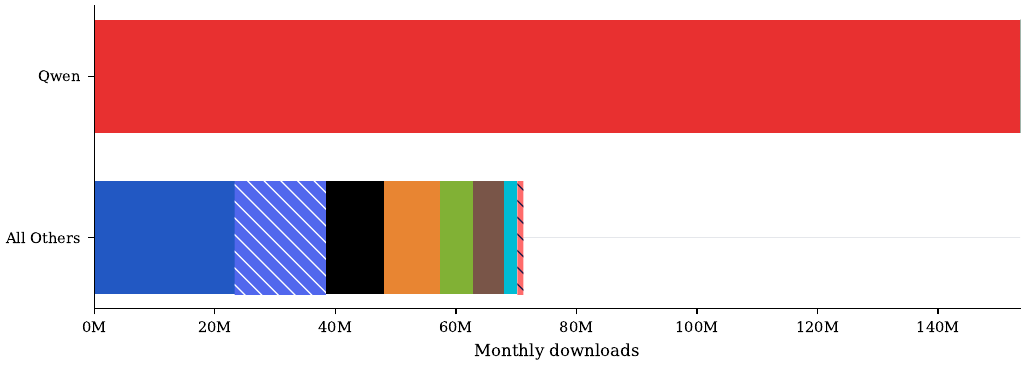}\\
  \cblock{121}{85}{72} Zhipu AI \quad \chblock{81}{103}{237} DeepSeek \quad \cblock{34}{88}{195} Meta \quad \chblockdark{255}{107}{107} MiniMax \quad \cblock{232}{133}{50} Mistral \quad \cblock{0}{188}{212} Moonshot AI \\
  \cblock{129}{177}{53} NVIDIA \quad \cblock{0}{0}{0} OpenAI \quad \cblock{232}{48}{48} Qwen
  \caption{Monthly Qwen downloads compared to combined downloads from all other major organizations. In February 2026, Qwen alone generated 153.6M downloads, more than double the combined 71.2M from the next eight major organizations.}
  \label{fig:qwen_ecosystem}
\end{figure}

\paragraph{DeepSeek and Qwen's complementary roles.}
Alibaba's Qwen and DeepSeek are the two most prominent open model families built in China, each having built critical acclaim through different styles of building and releasing models.
Qwen is the adoption leader by releasing numerous models across a range of sizes and abilities, where DeepSeek releases the most used, large MoE models.

\begin{figure}[H]
  \centering
  \includegraphics[width=0.9\linewidth]{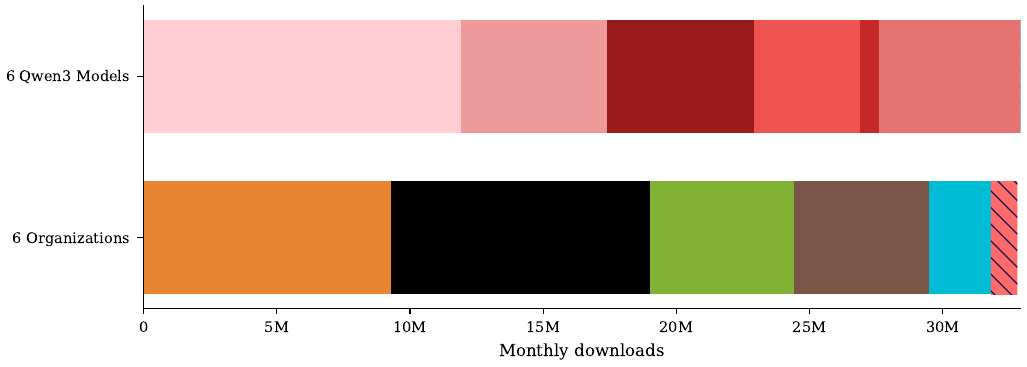}\\
  Qwen3: \ \ \cblock{255}{205}{210} 0.6B \quad \cblock{239}{154}{154} 1.7B \quad \cblock{229}{115}{115} 4B \quad \cblock{239}{83}{80} 4B Instruct \quad \cblock{198}{40}{40} 4B Thinking \quad \cblock{153}{27}{27} 8B \\
  \cblock{121}{85}{72} Zhipu AI \quad \chblockdark{255}{107}{107} MiniMax \quad \cblock{232}{133}{50} Mistral \quad \cblock{0}{188}{212} Moonshot AI \quad \cblock{129}{177}{53} NVIDIA \quad \cblock{0}{0}{0} OpenAI
  \caption{Monthly downloads for selected Qwen 3 models, including their quantized variants (0.6B--8B) versus competing organizations. 
  In February 2026, six Qwen 3 models combined for 32.9M downloads, roughly the same amount as the 32.8M generated by six competing organizations.}
  \label{fig:qwen3_effect}
\end{figure}

Figure~\ref{fig:qwen_ecosystem} shows that in February 2026, Qwen alone generated 153.6M monthly downloads -- more than double the combined 71.2M from the 8 other, leading open model builders.
This effect might be exaggerated in its magnitude as Qwen3.5~\citep{qwen2026qwen35} was released in February 2026, but the comparison remains accurate for other timeframes (e.g. Qwen had 87.5M downloads in December 2025, relative to the 61.3M of Meta Llama, DeepSeek, OpenAI, Mistral, Nvidia, Z.ai, Kimi, and MiniMax combined).
The driver of this gap is Qwen's smaller models -- just six small models of the Qwen3 series (out of 66 Qwen3 models in the complete family), i.e., Qwen3-0.6B -- Qwen3 8B have as many monthly downloads as six leading model organizations \textit{combined}, namely Zhipu AI, MiniMax, Mistral, Moonshot, NVIDIA and OpenAI (Figure~\ref{fig:qwen3_effect}).

\begin{figure}[H]
  \centering
  \includegraphics[width=0.9\linewidth]{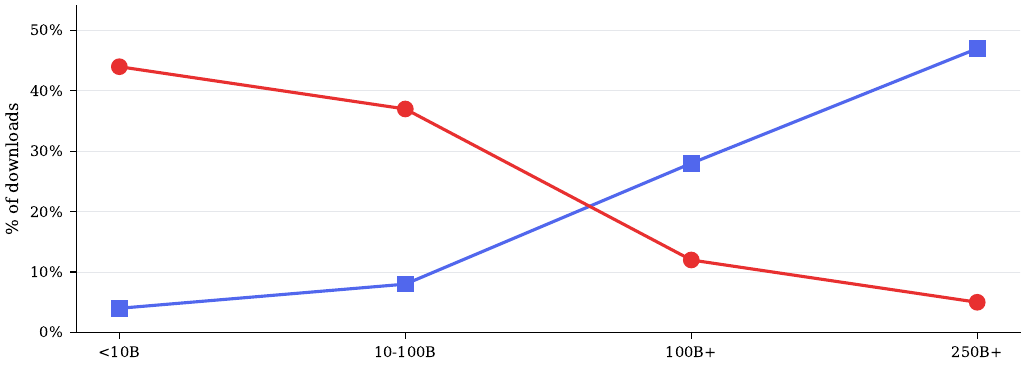}\\
  \ccircle{232}{48}{48} Qwen \quad \cblock{81}{103}{237} DeepSeek
  \caption{Download share by model size for DeepSeek vs.\ Qwen. DeepSeek dominates the 250B+ segment (47\%) while Qwen leads sub-10B (44\%), showing complementary specialization.}
  \label{fig:deepseek_size_lead}
\end{figure}

DeepSeek, by contrast, has a major adoption lead relative to Qwen in the lifetime use of the largest models.
Figure~\ref{fig:deepseek_size_lead} shows that DeepSeek captures 47\% of total tracked downloads in the 250B+ segment, while Qwen leads sub-10B models with 44\%.
These large model sizes represent the only model class where Qwen does not have a significant lead in adoption metrics.

\paragraph{Summary.}
Beyond Qwen's clear dominance in adoption metrics, the remaining lead model families each tell a distinct story:
\begin{itemize}
  \item \textbf{Mistral} was the earliest leader of the open language model ecosystem, highlighted by derivative share, peaking at 57\% in December 2023 on the strength of Mistral 7B~\citep{jiang2023mistral7b} and Mixtral~\citep{jiang2024mixtral}, but cumulative downloads of the models never accelerated to match Qwen or Llama's success.
  \item \textbf{DeepSeek} overtook Mistral in cumulative downloads in January 2026 (128.2M vs.\ 124.3M). Its impact is larger than downloads suggest: on OpenRouter, DeepSeek V3~\citep{deepseekai2024deepseekv3} and R1~\citep{deepseekai2025deepseekr1} accounted for up to 75.6\% of inference tokens in June 2025 and still held 31.1\% by January 2026 (Figure~\ref{fig:token_share_org}), reflecting heavy usage of a small number of flagship models rather than a broad derivative ecosystem.
  \item \textbf{OpenAI} is the most recent entrant, releasing GPT-OSS~\citep{openai2025gptoss} models starting in September 2025. By spring 2026, OpenAI's monthly downloads from a handful of models have surpassed those of Mistral's entire portfolio of historical models.
  \item \textbf{Meta} remains second in cumulative downloads (476.0M) but following the release of Llama 4 has collapsed on inference platforms and plateaued in adoption metrics, falling from a 37.4\% inference token share peak in January 2025 to zero by August 2025, replaced by a rotating cast of Chinese model providers.
\end{itemize}

\FloatBarrier

\subsection{New Entrants (Nvidia, Moonshot AI, MiniMax, Z.ai, ...)}

Crucially, many of the labs known for their frontier-level, large MoE models, such as MiniMax, Moonshot AI, or Z.ai, are far from dominating the adoption metrics of open models.
These organizations are competing at a fraction of the adoption level of the leaders, but they have meaningful growth that is worth watching.

\begin{figure}[H]
  \centering
  \includegraphics[width=\linewidth]{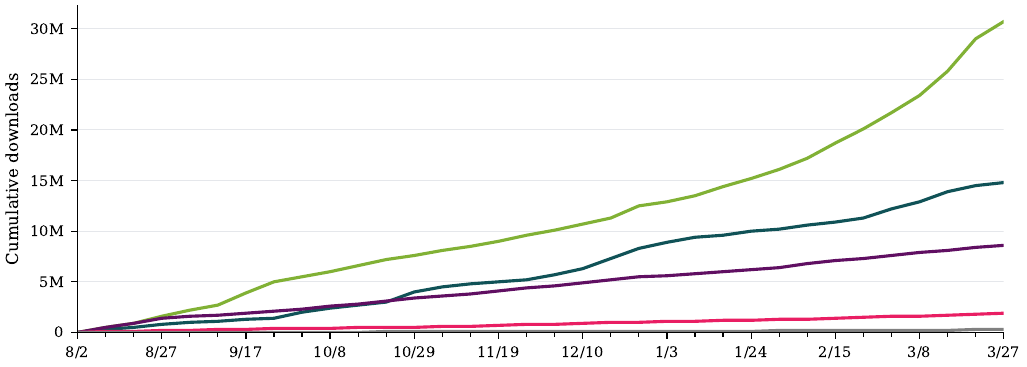}\\
  \cblock{233}{30}{99} AI21 \quad \cblock{16}{82}{87} Ai2 \quad \cblock{128}{128}{128} Arcee \quad \cblock{96}{15}{98} IBM \quad \cblock{129}{177}{53} NVIDIA
  \caption{Cumulative downloads from new American open model entrants since August 2025. By March 27, 2026, NVIDIA reached 30.7M downloads, Ai2 14.8M, and IBM 8.6M, still far behind Qwen's 942.1M cumulative downloads.}
  \label{fig:us_entrants}
\end{figure}

The summer of 2025 was characterized by a wave of very capable open models from a variety of Chinese organizations.
This series of impressive, Chinese models prompted the writing of the original The ATOM Project memo in August 2025.
Since, we tracked American labs as they attempted to catch up with more useful open models.
Figure~\ref{fig:us_entrants} tracks the newer and smaller American entrants via cumulative downloads since August 2025: NVIDIA leads with 30.7M cumulative downloads driven by its Nemotron family~\citep{nvidia2024nemotron4,nvidia2025nemotron3nano}, followed by the Allen Institute for AI (Ai2) at 14.8M with OLMo~\citep{olmo2025olmo3} and IBM at 8.6M with Granite~\citep{granite2024language}.
The scale gap remains large -- all US entrants combined account for roughly 56M downloads versus Qwen's 942.1M -- but their growth trajectories show sustained progress.

\begin{figure}[H]
  \centering
  \includegraphics[width=\linewidth]{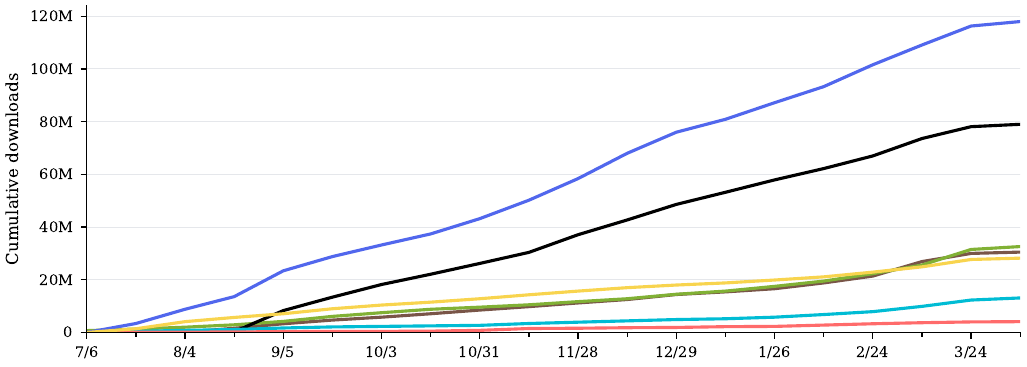}\\
  \cblock{121}{85}{72} Zhipu AI \quad \cblock{81}{103}{237} DeepSeek \quad \cblock{248}{212}{78} Hugging Face \quad \cblock{255}{107}{107} MiniMax \quad \cblock{0}{188}{212} Moonshot AI \quad \cblock{129}{177}{53} NVIDIA \quad \cblock{0}{0}{0} OpenAI
  \caption{Cumulative downloads since July 2025 for organizations competing with DeepSeek. DeepSeek leads with 118.0M downloads, ahead of OpenAI at 79.0M.}
  \label{fig:taking_on_deepseek}
\end{figure}

Figure~\ref{fig:taking_on_deepseek} captures a broader set of organizations competing with DeepSeek, OpenAI, and Mistral as potential new organizations with top 5 overall downloads, plotted since July 2025, when we expanded our data collection tooling to encompass every organization.
Examples here include Hugging Face (SmolLM), MiniMax (M Series), and Moonshot AI (Kimi models).

Inference data from OpenRouter reveals additional entrants not captured in download metrics: Xiaomi's MiMo-V2-Flash~\citep{xiaomi2026mimov2flash}, a 309B-parameter MoE released in December 2025, surged from zero to 27.2\% of inference token share by January 2026, illustrating how quickly popular models can appear on inference platforms without a corresponding footprint in Hugging Face downloads.

\clearpage
\section{The Relative Adoption Metric (RAM)}
\label{sec:ram-method}

\subsection{Method}
Standard practice for measuring the impact of open models is to compare total and monthly download numbers to those of other models.
These rough metrics don't account for the variable levels of adoption across model size categories and relative adoption of a new model as the overall adoption of open models grows rapidly.
We set out to design a new adoption metric, building on a finer-grained version of the same download data, to better assess models across size categories and time.

For example, it is intuitively simple that a 1.5B model routinely accumulates 10--50$\times$ more downloads than a 400B model simply because it is cheaper to run, easier to integrate into CI/CD pipelines, and more frequently loaded in automated testing.
We introduce the \textbf{Relative Adoption Metric (RAM)} to normalize adoption trajectories within size cohorts.
RAM is particularly useful for medium-to-large models, where download numbers can be more precisely contextualized against a small set of peer models.

\paragraph{Definition.}
For a given model $m$ in size bucket $b$ (same as in Sec.~\ref{sec:categorization}, with more information in Fig.~\ref{fig:downloads_by_size}), at milestone $t$ days post-release:
\begin{equation}
  \text{RAM}(m, t) = \frac{D(m, t)}{C_{10}(b, t)}
\end{equation}
where $D(m,t)$ is the cumulative download count for model $m$ at $t$ days after release, and $C_{10}(b,t)$ is the top-10 cutoff for bucket $b$ at the same milestone (the top models used for this analysis are documented in Appendix~\ref{sec:appendix-ram-top-models}).
A RAM score above 1.0$\times$ means the model is ahead of its size-class target.

\paragraph{Milestones.}
Scores are computed at seven fixed post-release milestones: 7, 14, 30, 60, 90, 180, and 365 days.
Early milestones (7--30 days) capture the initial launch momentum, while later milestones reflect sustained community interest and adoption.

\begin{figure}[H]
  \centering
  \includegraphics[width=\linewidth]{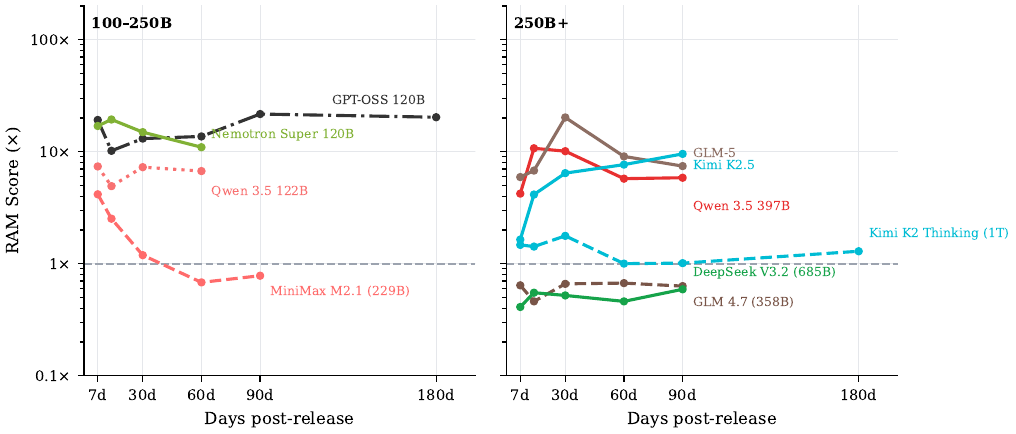}
  \caption{RAM trajectories for notable recent model releases. Each line shows cumulative downloads relative to the monotonic top-10 cutoff for that size bucket. Where official Hugging Face releases include multiple published weight-format variants, downloads are aggregated across those variants. RAM baseline: 2026-Q2.}
  \label{fig:ram_case_studies}
\end{figure}

\paragraph{Statistical Design.}
The reference set for each of the seven size buckets starts from the reviewed top-20 model candidate pool in the 2026-Q2 RAM baseline.
At each time checkpoint, the RAM denominator is the 10th-highest release-aligned download count among candidates with enough history.
This choice is motivated by extreme right-skew of a few models with drastically higher downloads: in the 1--5B bucket, a single breakout model can pull the mean upward by roughly an order of magnitude relative to the median.
Using the 10th top download number per time checkpoint produces reference values that are robust to such outliers and that give an intuitive baseline for interpreting new model adoption.
We make these reference values monotonic by carrying forward the previous cutoff when a later raw cutoff is lower.
Full reference statistics are provided in the supplementary materials; representative reference curves are shown in Figure~\ref{fig:ram_reference}, and case studies for recent models are presented in Figure~\ref{fig:ram_case_studies}.

Each RAM score is tied to a specific snapshot corresponding to when the candidate pools and top-10 cutoffs were pulled.
For our analysis, we update the RAM benchmarks quarterly and are constructing an API of model adoption for use by more researchers.

Additional details and data on RAM are included in Appendix~\ref{appd:ram}.

\subsection{Using RAM to Measure New Models}
\label{sec:ram-scores}


RAM shows a clear ranking of recent models over time within as little as 30 days following the model release.
We tested RAM on a variety of open models released in the last 6 months to showcase models that are over- or under-performing on adoption metrics relative to community excitement.

For example, using the 2026-Q2 RAM baseline, the RAM framework makes the February 2026 Qwen3.5 rollout easier to compare across scales.
In the 1--5B bucket, shown in Fig.~\ref{fig:ram_case_studies_small}, \texttt{Qwen3.5-4B} reached 6.97$\times$, 8.93$\times$, 4.80$\times$, and 4.09$\times$ the bucket's top-10 cutoff at 7, 14, 30, and 60 days.
This is a strong launch, but not a category outlier because small-model baselines are already high: the 1--5B top-10 cutoff was 24K downloads at 7 days, 84K at 14 days, 500K at 30 days, and 1.5M at 60 days.
For comparison, DeepSeek OCR (3B) still launched higher within the same bucket, reaching 12.68$\times$ and 15.45$\times$ at 7 and 14 days.
The clearest breakout model was \texttt{Qwen3.5-35B-A3B}, which reached 10.44$\times$ at 7 days, 12.13$\times$ at 14 days, 8.01$\times$ at 30 days, and 8.75$\times$ at 60 days, placing it among the hottest launches in our reference set.
In the 100--250B bucket, shown in Fig.~\ref{fig:ram_case_studies}, \texttt{Qwen3.5-122B-A10B} reached 7.38$\times$, 4.92$\times$, 7.27$\times$, and 6.71$\times$ at 7, 14, 30, and 60 days.
At the giant end, \texttt{Qwen3.5-397B-A17B} reached 4.22$\times$, 10.72$\times$, 10.10$\times$, and 5.85$\times$ at 7, 14, 30, and 90 days.

Outside of Qwen~3.5 and the aforementioned models, there are clear leading and lagging models released in recent months.
In 100--250B, GPT-OSS 120B and Nemotron Super 120B are clear outliers following their launches, approaching levels to be some of the most-downloaded models of all time: GPT-OSS reached 19.19$\times$ at 7 days and was still at 20.31$\times$ at 180 days, while Nemotron Super 120B opened at 16.96$\times$ and 19.38$\times$ at 7 and 14 days and remained at 10.96$\times$ at 60 days.
A counter example is MiniMax M2.1, which launched strongly but faded back below the top-10 boundary, moving from 4.16$\times$ at 7 days to 0.78$\times$ at 90 days.

In 250B+, GLM-5 was a strong recent launch, reaching 5.93$\times$, 6.78$\times$, and 20.21$\times$ scores at 7, 14, and 30 days and remaining at 7.43$\times$ at 90 days, while GLM 4.7 and DeepSeek V3.2 stayed below the top-10 boundary throughout their post-release monitoring window.
Other models include Kimi K2 Thinking, which mostly tracked around the top-10 boundary, or Kimi K2.5, which accelerated from 1.64$\times$ at 7 days to 9.55$\times$ at 90 days.

Further data for this section is included in Appendix~\ref{appd:ram} and specifically Table~\ref{tab:ram_scores}.

\FloatBarrier

\section{Conclusion}
\label{sec:conclusion}

The ATOM Report is a first step in creating a focused set of tools for understanding the relative adoption of the leading open language models.
Core to the methodology is a goal to bring both finer grained measurement and historical analysis to an area that traditionally has been defined by very crude and noisy measurements.
The data reinforces many key trends across the ecosystem, from Qwen's growing lead to Llama's stagnation and the role of many new entrants.
Crucially, this data is all one small step towards a clearer picture, and the ecosystem needs to continue to refine its data sharing and transparency, in order to better showcase potential new entrants on the value of releasing a new type of open-weights model.

\paragraph{Acknowledgments.}
We thank Hugging Face and OpenRouter for sharing private data that made this analysis possible, and Artificial Analysis, Arena, and Epoch AI for making valuable data publicly available.
Thank you to Caithrin Rintoul for invaluable support for this project.

\bibliographystyle{plainnat-etal}
\bibliography{references}

\clearpage
\appendix
\section{Related Work}
\label{sec:related_work}

A handful of other groups have done work collecting and analyzing subsets of the data we discuss in this report. 
In the closest related work, \citet{longpre2025economiesopenintelligence} measure the participation dynamics across the entire ecosystem, i.e. not solely focused on key language models since ChatGPT, and reaches many overlapping conclusions as this report.\footnote{Also see related, recurring work in this direction from Hugging Face directly~\citep{hf_state_of_os_spring26}.}
The AI Index provides annual ecosystem-level indicators for releases, capabilities, and deployment trends \citep{maslej2024aiindex,maslej2025aiindex}. 
Transparency-centered efforts evaluate disclosure practices rather than adoption outcomes, including the Foundation Model Transparency Index and related reporting frameworks \citep{bommasani2023fmti,bommasani2024fmti,wan2025fmti,bommasani2024transparencyreports}. 
Platform-focused studies of Hugging Face characterize repository growth, maintenance behavior, and data quality of the platform itself \citep{ait2023hfhub_empirical,castano2023hfhub_models}.
Governance and data-provenance work adds complementary evidence on licensing, attribution, and the contraction of the open data commons \citep{longpre2023dataprovenance,longpre2024consentincrisis}. 




\section{Q2 2026 Top Downloaded Models by Category}
\label{sec:appendix-ram-top-models}

We include the top 10 downloaded models per size category from the 2026-Q2 RAM candidate pools, as documented in Sec.~\ref{sec:ram-method}.
The production API uses a reviewed top-20 candidate pool and computes the 10th-place cutoff separately for each days-since-release checkpoint.
Download counts in this table are from the 2026-Q2 RAM baseline snapshot on May 23, 2026.

\begin{small}
\setlength{\tabcolsep}{8pt}
\renewcommand{\arraystretch}{1.2}

\subsection*{Under 1B}
\begin{tabular}{@{}p{0.48\linewidth}p{0.48\linewidth}@{}}
1. \href{https://huggingface.co/Qwen/Qwen3-0.6B}{Qwen3-0.6B} (108M) & 6. \href{https://huggingface.co/google/t5gemma-b-b-prefixlm}{t5gemma-b-b-prefixlm} (17.5M) \\
2. \href{https://huggingface.co/Qwen/Qwen2.5-0.5B-Instruct}{Qwen2.5-0.5B-Instruct} (42.9M) & 7. \href{https://huggingface.co/Qwen/Qwen2.5-Coder-0.5B-Instruct}{Qwen2.5-Coder-0.5B-Instruct} (13.9M) \\
3. \href{https://huggingface.co/google/gemma-3-1b-it}{gemma-3-1b-it} (36.5M) & 8. \href{https://huggingface.co/EleutherAI/pythia-160m}{pythia-160m} (13.4M) \\
4. \href{https://huggingface.co/Qwen/Qwen2.5-0.5B}{Qwen2.5-0.5B} (21.1M) & 9. \href{https://huggingface.co/HuggingFaceTB/SmolLM2-135M}{SmolLM2-135M} (13M) \\
5. \href{https://huggingface.co/microsoft/Florence-2-large}{Florence-2-large} (20.8M) & 10. \href{https://huggingface.co/microsoft/Florence-2-base}{Florence-2-base} (11.8M) \\
\end{tabular}

\subsection*{1-5B}
\begin{tabular}{@{}p{0.48\linewidth}p{0.48\linewidth}@{}}
1. \href{https://huggingface.co/Qwen/Qwen3-VL-2B-Instruct}{Qwen3-VL-2B-Instruct} (206.8M) & 6. \href{https://huggingface.co/meta-llama/Llama-3.2-1B}{Llama-3.2-1B} (52.4M) \\
2. \href{https://huggingface.co/Qwen/Qwen2.5-1.5B-Instruct}{Qwen2.5-1.5B-Instruct} (175M) & 7. \href{https://huggingface.co/Qwen/Qwen3-4B}{Qwen3-4B} (46.3M) \\
3. \href{https://huggingface.co/Qwen/Qwen2.5-3B-Instruct}{Qwen2.5-3B-Instruct} (89.4M) & 8. \href{https://huggingface.co/Qwen/Qwen3-4B-Instruct-2507}{Qwen3-4B-Instruct-2507} (44.5M) \\
4. \href{https://huggingface.co/Qwen/Qwen2.5-VL-3B-Instruct}{Qwen2.5-VL-3B-Instruct} (80M) & 9. \href{https://huggingface.co/meta-llama/Llama-3.2-3B-Instruct}{Llama-3.2-3B-Instruct} (42M) \\
5. \href{https://huggingface.co/meta-llama/Llama-3.2-1B-Instruct}{Llama-3.2-1B-Instruct} (68.6M) & 10. \href{https://huggingface.co/Qwen/Qwen3-1.7B}{Qwen3-1.7B} (39.4M) \\
\end{tabular}

\subsection*{7-9B}
\begin{tabular}{@{}p{0.48\linewidth}p{0.48\linewidth}@{}}
1. \href{https://huggingface.co/meta-llama/Llama-3.1-8B-Instruct}{Llama-3.1-8B-Instruct} (152.4M) & 6. \href{https://huggingface.co/meta-llama/Meta-Llama-3-8B}{Meta-Llama-3-8B} (42.3M) \\
2. \href{https://huggingface.co/Qwen/Qwen2.5-7B-Instruct}{Qwen2.5-7B-Instruct} (133.8M) & 7. \href{https://huggingface.co/meta-llama/Meta-Llama-3-8B-Instruct}{Meta-Llama-3-8B-Instruct} (42M) \\
3. \href{https://huggingface.co/Qwen/Qwen2.5-VL-7B-Instruct}{Qwen2.5-VL-7B-Instruct} (64.6M) & 8. \href{https://huggingface.co/mistralai/Mistral-7B-Instruct-v0.3}{Mistral-7B-Instruct-v0.3} (31.3M) \\
4. \href{https://huggingface.co/Qwen/Qwen3-8B}{Qwen3-8B} (63M) & 9. \href{https://huggingface.co/meta-llama/Llama-2-7b-hf}{Llama-2-7b-hf} (31.2M) \\
5. \href{https://huggingface.co/mistralai/Mistral-7B-Instruct-v0.2}{Mistral-7B-Instruct-v0.2} (58.7M) & 10. \href{https://huggingface.co/Qwen/Qwen2-VL-7B-Instruct}{Qwen2-VL-7B-Instruct} (30.2M) \\
\end{tabular}

\subsection*{10-50B}
\begin{tabular}{@{}p{0.48\linewidth}p{0.48\linewidth}@{}}
1. \href{https://huggingface.co/openai/gpt-oss-20b}{gpt-oss-20b} (68.3M) & 6. \href{https://huggingface.co/mistralai/Mixtral-8x7B-Instruct-v0.1}{Mixtral-8x7B-Instruct-v0.1} (21.2M) \\
2. \href{https://huggingface.co/Qwen/Qwen2.5-14B-Instruct}{Qwen2.5-14B-Instruct} (37.6M) & 7. \href{https://huggingface.co/meta-llama/Llama-3.2-11B-Vision-Instruct}{Llama-3.2-11B-Vision-Instruct} (17.9M) \\
3. \href{https://huggingface.co/Qwen/Qwen3-32B}{Qwen3-32B} (34.1M) & 8. \href{https://huggingface.co/google/gemma-3-12b-it}{gemma-3-12b-it} (17.2M) \\
4. \href{https://huggingface.co/deepseek-ai/DeepSeek-R1-Distill-Qwen-32B}{DeepSeek-R1-Distill-Qwen-32B} (24.7M) & 9. \href{https://huggingface.co/Qwen/Qwen3-14B}{Qwen3-14B} (17.1M) \\
5. \href{https://huggingface.co/Qwen/Qwen2.5-32B-Instruct}{Qwen2.5-32B-Instruct} (22.2M) & 10. \href{https://huggingface.co/meta-llama/Llama-2-13b-chat-hf}{Llama-2-13b-chat-hf} (15.5M) \\
\end{tabular}

\subsection*{50-100B}
\begin{tabular}{@{}p{0.48\linewidth}p{0.48\linewidth}@{}}
1. \href{https://huggingface.co/meta-llama/Llama-3.1-70B-Instruct}{Llama-3.1-70B-Instruct} (21.6M) & 6. \href{https://huggingface.co/meta-llama/Meta-Llama-3-70B-Instruct}{Meta-Llama-3-70B-Instruct} (6.1M) \\
2. \href{https://huggingface.co/Qwen/Qwen3-Next-80B-A3B-Instruct}{Qwen3-Next-80B-A3B-Instruct} (15.3M) & 7. \href{https://huggingface.co/Qwen/Qwen2.5-VL-72B-Instruct}{Qwen2.5-VL-72B-Instruct} (6.1M) \\
3. \href{https://huggingface.co/meta-llama/Llama-3.3-70B-Instruct}{Llama-3.3-70B-Instruct} (11.9M) & 8. \href{https://huggingface.co/meta-llama/Llama-2-70b-chat-hf}{Llama-2-70b-chat-hf} (4.9M) \\
4. \href{https://huggingface.co/Qwen/Qwen2.5-72B-Instruct}{Qwen2.5-72B-Instruct} (6.7M) & 9. \href{https://huggingface.co/deepseek-ai/DeepSeek-R1-Distill-Llama-70B}{DeepSeek-R1-Distill-Llama-70B} (4.6M) \\
5. \href{https://huggingface.co/OpenGVLab/InternVL3-78B}{InternVL3-78B} (6.3M) & 10. \href{https://huggingface.co/meta-llama/Meta-Llama-3-70B}{Meta-Llama-3-70B} (3.4M) \\
\end{tabular}

\subsection*{100-250B}
\begin{tabular}{@{}p{0.48\linewidth}p{0.48\linewidth}@{}}
1. \href{https://huggingface.co/openai/gpt-oss-120b}{gpt-oss-120b} (37.8M) & 6. \href{https://huggingface.co/Qwen/Qwen3-235B-A22B}{Qwen3-235B-A22B} (4.4M) \\
2. \href{https://huggingface.co/mistralai/Mixtral-8x22B-Instruct-v0.1}{Mixtral-8x22B-Instruct-v0.1} (6.1M) & 7. \href{https://huggingface.co/OpenGVLab/InternVL3_5-241B-A28B-Instruct}{InternVL3\_5-241B-A28B-Instruct} (4.1M) \\
3. \href{https://huggingface.co/mistralai/Mistral-Large-Instruct-2407}{Mistral-Large-Instruct-2407} (5M) & 8. \href{https://huggingface.co/Qwen/Qwen3-VL-235B-A22B-Instruct}{Qwen3-VL-235B-A22B-Instruct} (4M) \\
4. \href{https://huggingface.co/mistralai/Mistral-Large-Instruct-2411}{Mistral-Large-Instruct-2411} (4.9M) & 9. \href{https://huggingface.co/Qwen/Qwen3-235B-A22B-Instruct-2507-FP8}{Qwen3-235B-A22B-Instruct-2507-FP8} (3.5M) \\
5. \href{https://huggingface.co/mistralai/Mixtral-8x22B-v0.1}{Mixtral-8x22B-v0.1} (4.8M) & 10. \href{https://huggingface.co/Qwen/Qwen3-VL-235B-A22B-Thinking}{Qwen3-VL-235B-A22B-Thinking} (3.5M) \\
\end{tabular}

\subsection*{250B+}
\begin{tabular}{@{}p{0.48\linewidth}p{0.48\linewidth}@{}}
1. \href{https://huggingface.co/deepseek-ai/DeepSeek-R1}{DeepSeek-R1} (24.8M) & 6. \href{https://huggingface.co/deepseek-ai/DeepSeek-R1-0528}{DeepSeek-R1-0528} (8.8M) \\
2. \href{https://huggingface.co/meta-llama/Llama-3.1-405B}{Llama-3.1-405B} (20.6M) & 7. \href{https://huggingface.co/zai-org/GLM-5-FP8}{GLM-5-FP8} (7.9M) \\
3. \href{https://huggingface.co/deepseek-ai/DeepSeek-V3}{DeepSeek-V3} (16.5M) & 8. \href{https://huggingface.co/deepseek-ai/DeepSeek-V3-0324}{DeepSeek-V3-0324} (5.2M) \\
4. \href{https://huggingface.co/deepseek-ai/DeepSeek-V3.2}{DeepSeek-V3.2} (16M) & 9. \href{https://huggingface.co/meta-llama/Llama-3.1-405B-Instruct}{Llama-3.1-405B-Instruct} (3.9M) \\
5. \href{https://huggingface.co/moonshotai/Kimi-K2.5}{Kimi-K2.5} (11.7M) & 10. \href{https://huggingface.co/Qwen/Qwen3.5-397B-A17B}{Qwen3.5-397B-A17B} (3.8M) \\
\end{tabular}

\end{small}

\section{Additional RAM Details}
\label{appd:ram}

\begin{figure}[H]
  \centering
  \includegraphics[width=\linewidth]{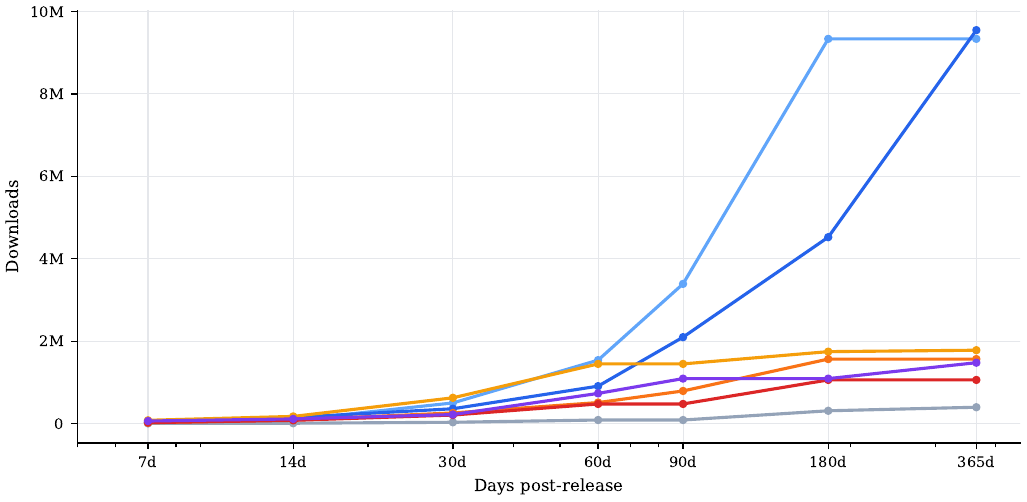}\\
  \cblock{148}{163}{184} {$<$}1B \quad \cblock{96}{165}{250} 1--5B \quad \cblock{37}{99}{235} 7--9B \quad \cblock{245}{158}{11} 10--50B \quad \cblock{249}{115}{22} 50--100B \quad \cblock{220}{38}{38} 100--250B \quad \cblock{124}{58}{237} 250B+
  \caption{RAM reference curves showing the monotonic top-10 cutoff cumulative downloads for each size category over time.}
  \label{fig:ram_reference}
\end{figure}

The top-10 boundary download counts over time, used to compute the RAM scores, are shown in Figure~\ref{fig:ram_reference}.
The plotted baseline is monotonic: if a later raw cutoff is lower, the previous higher cutoff is carried forward.
As our data backlog grows, this carry-forward rule should become unnecessary.
The smallest models have larger outliers, as shown in Appendix~\ref{sec:appendix-ram-top-models}, where models such as Qwen3-0.6B with 108M downloads, Qwen3-VL-2B-Instruct with 206.8M, or Llama-3.1-8B-Instruct with 152.4M have substantially more downloads than the 10th-ranked model in those categories (Florence-2-base with 11.8M, Qwen3-1.7B with 39.4M, and Qwen2-VL-7B-Instruct with 30.2M).

\begin{figure}[H]
  \centering
  \includegraphics[width=\linewidth]{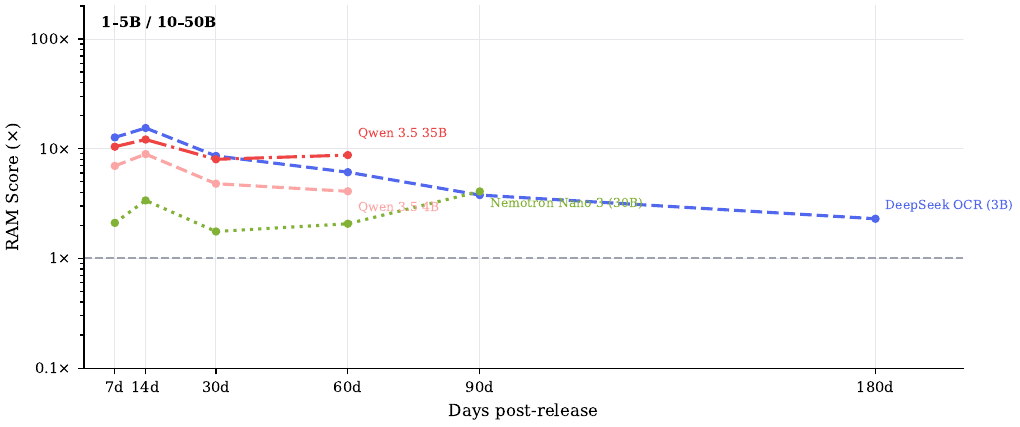}
  \caption{RAM trajectories for small and medium models (1--5B and 10--50B buckets). DeepSeek OCR (3B) remained above the top-10 boundary through 180 days, while Qwen 3.5 35B stayed at 8.75$\times$ at 60 days. RAM baseline: 2026-Q2.}
  \label{fig:ram_case_studies_small}
\end{figure}

Figure~\ref{fig:ram_case_studies_small} shows RAM trajectories for a few small and medium models (1--5B and 10--50B), complementing the 100--250B and 250B+ panels in the main text.
Table~\ref{tab:ram_scores} reports exact RAM scores and cumulative downloads at each milestone for all case-study models.

\begin{table}[H]
\centering
\small
\caption{RAM scores (top row, $\times$ top-10 cutoff) and cumulative downloads (bottom row) for a set of case-study models at each post-release milestone. Empty cells indicate the model has not yet reached that milestone. RAM baseline: 2026-Q2.}
\label{tab:ram_scores}
\begin{tabular}{@{}clrrrrrr@{}}
\toprule
Bucket & Model & 7d & 14d & 30d & 60d & 90d & 180d \\
\midrule
\multirow{4}{*}{\textbf{1--5B}}
  & \multirow{2}{*}{DeepSeek OCR (3B)}
    & 12.68$\times$ & 15.45$\times$ & 8.59$\times$ & 6.11$\times$ & 3.78$\times$ & 2.30$\times$ \\
  & & 302K & 1.3M & 4.3M & 9.4M & 12.8M & 21.4M \\[2pt]
  & \multirow{2}{*}{Qwen 3.5 4B}
    & 6.97$\times$ & 8.93$\times$ & 4.80$\times$ & 4.09$\times$ & & \\
  & & 166K & 751K & 2.4M & 6.3M & & \\
\midrule
\multirow{4}{*}{\textbf{10--50B}}
  & \multirow{2}{*}{Qwen 3.5 35B}
    & 10.44$\times$ & 12.13$\times$ & 8.01$\times$ & 8.75$\times$ & & \\
  & & 822K & 2.1M & 5.0M & 12.7M & & \\[2pt]
  & \multirow{2}{*}{Nemotron Nano 3 (30B)}
    & 2.11$\times$ & 3.38$\times$ & 1.76$\times$ & 2.07$\times$ & 4.08$\times$ & \\
  & & 166K & 585K & 1.1M & 3.0M & 5.9M & \\
\midrule
\multirow{8}{*}{\textbf{100--250B}}
  & \multirow{2}{*}{GPT-OSS 120B}
    & 19.19$\times$ & 10.18$\times$ & 13.09$\times$ & 13.68$\times$ & 21.68$\times$ & 20.31$\times$ \\
  & & 429K & 788K & 2.7M & 6.5M & 10.3M & 21.5M \\[2pt]
  & \multirow{2}{*}{Nemotron Super 120B}
    & 16.96$\times$ & 19.38$\times$ & 14.94$\times$ & 10.96$\times$ & & \\
  & & 379K & 1.5M & 3.1M & 5.2M & & \\[2pt]
  & \multirow{2}{*}{Qwen 3.5 122B}
    & 7.38$\times$ & 4.92$\times$ & 7.27$\times$ & 6.71$\times$ & & \\
  & & 165K & 381K & 1.5M & 3.2M & & \\[2pt]
  & \multirow{2}{*}{MiniMax M2.1 (229B)}
    & 4.16$\times$ & 2.52$\times$ & 1.19$\times$ & 0.68$\times$ & 0.78$\times$ & \\
  & & 93K & 195K & 246K & 323K & 372K & \\
\midrule
\multirow{12}{*}{\textbf{250B+}}
  & \multirow{2}{*}{GLM-5}
    & 5.93$\times$ & 6.78$\times$ & 20.21$\times$ & 9.07$\times$ & 7.43$\times$ & \\
  & & 362K & 759K & 4.4M & 6.6M & 8.1M & \\[2pt]
  & \multirow{2}{*}{Qwen 3.5 397B}
    & 4.22$\times$ & 10.72$\times$ & 10.10$\times$ & 5.74$\times$ & 5.85$\times$ & \\
  & & 258K & 1.2M & 2.2M & 4.2M & 6.4M & \\[2pt]
  & \multirow{2}{*}{Kimi K2.5}
    & 1.64$\times$ & 4.14$\times$ & 6.43$\times$ & 7.66$\times$ & 9.55$\times$ & \\
  & & 100K & 463K & 1.4M & 5.6M & 10.4M & \\[2pt]
  & \multirow{2}{*}{Kimi K2 Thinking (1T)}
    & 1.47$\times$ & 1.42$\times$ & 1.77$\times$ & 1.00$\times$ & 1.01$\times$ & 1.29$\times$ \\
  & & 90K & 159K & 385K & 732K & 1.1M & 1.4M \\[2pt]
  & \multirow{2}{*}{GLM 4.7 (358B)}
    & 0.64$\times$ & 0.46$\times$ & 0.66$\times$ & 0.67$\times$ & 0.63$\times$ & \\
  & & 39K & 51K & 144K & 491K & 685K & \\[2pt]
  & \multirow{2}{*}{DeepSeek V3.2 (685B)}
    & 0.41$\times$ & 0.55$\times$ & 0.52$\times$ & 0.46$\times$ & 0.59$\times$ & \\
  & & 25K & 61K & 114K & 339K & 648K & \\
\bottomrule
\end{tabular}
\end{table}

\end{document}